\documentstyle[sprocl,epsfig]{article}

\def\be{\begin{equation}}
\def\ee{\end{equation}}
\def\bea{\begin{eqnarray}}
\def\eea{\end{eqnarray}}
\def\mr{\mathrm}

\begin{document}

\title{TOWARDS A FRAGMENTATION MODEL FOR SHERPA
  \footnote{Presented by S. Schumann at the {\it International Conference on Linear Colliders}, 
    19-23 April 2004, Paris, France.}\\}

\author{T.\ GLEISBERG, S.\ H{\"O}CHE, F.\ KRAUSS, A.\ SCH{\"A}LICKE,
        S.\ SCHUMANN, \\ J.\ WINTER, G.\ SOFF}

\address{Institut f\"ur theoretische Physik, TU Dresden, D-01062 Dresden, 
  Germany\\E-mail: steffen@theory.phy.tu-dresden.de}

\maketitle\abstracts{
Some results highlighting the status of a new version of a cluster fragmentation model 
for the Monte Carlo event generator Sherpa are presented. In its present version this 
model is capable of simulating $e^+e^-$ annihilation events into light-quark and gluon 
jets. We compare results for different multiplicity and momentum distributions to 
available SLD and LEP data as well as to results obtained with Herwig and Pythia.
}
  

\section{Introduction}
In view of the LHC starting to provide $pp$ collisions at $\sqrt{s}= 14$ 
TeV in 2007 and a future $e^+e^-$ linear collider in the TeV range, new 
tools for the Monte Carlo simulation of multi-hadron final states are 
required. These tools will once be faced with the most precise measurements 
in high energy particle physics -- this aspect dictating their improved physics 
content. In addition transparency and maintenance of these codes become an 
issue and therefore the object oriented language C++ was chosen to write them. 
Beside the rewrites of the well-established tools Pythia 
\cite{Sjostrand:2000wi} and Herwig 
\cite{Corcella:2000bw_and_2002jc}, namely Pythia7 \cite{Bertini:2000uh} 
and Herwig++ \cite{Gieseke:2004ag_Gieseke:2003hm}, another approach is 
available with the program Sherpa \cite{Gleisberg:2003xi}. One of the striking 
features of Sherpa is the inclusion of the CKKW prescription to combine multi-jet 
matrix elements with parton showers \cite{Catani:2001cc_Krauss:2002up}. This 
method allows a consistent description of multi-jet final states and a combination 
of such higher order calculations with the non-perturbative regime of hadron production 
in an universal manner. Even though the public Sherpa version \cite{sherpa} contains 
an interface to the Pythia string fragmentation, strong efforts have been made to develop 
a fragmentation model for Sherpa \cite{Winter:2003tt} that relies on the cluster 
fragmentation ansatz used within Herwig \cite{Webber:1983if}.


\section{The Cluster Model}
A typical cluster fragmentation model consists of two parts: first, primary clusters are formed
and, following this, such colour neutral states decay into hadrons and/or secondary clusters,
which, in turn, decay further. 

In the first step the partons emerging from the parton shower are brought to their constituent 
masses; this includes a finite gluon mass. The gluon then is forced to split into a light 
quark-antiquark ($q\bar q$) or antidiquark-diquark ($\bar DD$) pair. The resulting triplet and 
antitriplet states are combined to obtain colour-neutral clusters. The model thereby allows 
for the incorporation of soft colour reconnection effects, which lead to configurations 
that are beyond the planar structure given by the parton shower evolution. 
However, these configurations are suppressed by a combined weight of $1/N_C^2$ and a 
kinematic function. After this first step four different cluster types can arise, 
mesonic ($q\bar q$ and $\bar DD$), baryonic ($qD$)and antibaryonic 
($\bar D\bar q$).  

In the second step the primary clusters, continuously distributed in mass with a 
peak at low cluster masses, have to be transformed into observable hadrons featuring 
a discrete mass spectrum. This is achieved by binary cluster decays and converting 
individual clusters into single primary hadrons. The model so far does not incorporate 
the subsequent decay of unstable hadrons. This task is still handled by the corresponding 
Pythia routines. Beside the request of locality and low momentum transfer in cluster 
decays, the model relies on a dynamic separation of clusters and hadrons. This implies 
that according to the flavour of its constituents a cluster is supposed to be a hadron, 
if its mass is below a certain threshold. Similar to the case of cluster formation, the 
model for cluster decays incorporates the possibility for soft colour reconnection. 
Thereby a quark-antiquark pair  or antidiquark-diquark pair produced to disintegrate 
the cluster can recombine. Again these configurations are suppressed according to 
$1/N_C^2$ and a kinematic weight. 

\begin{figure}[ht!] 
  \begin{center}
    \epsfig{file=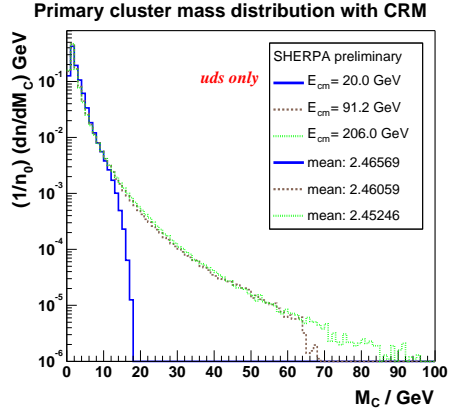,width=5.5cm,angle=0}
  \end{center}
  \vspace{-1mm}
  \caption{The primary cluster mass distribution in $e^+e^-$ annihilation events 
for different centre of mass energies using the colour reconnection model (CRM).} 
  \label{fig:clumass}
\end{figure}

\begin{figure}[hb!]
  \begin{center}
    \vspace*{-2mm}
    \begin{tabular}{cc}
      \mbox{\epsfig{file=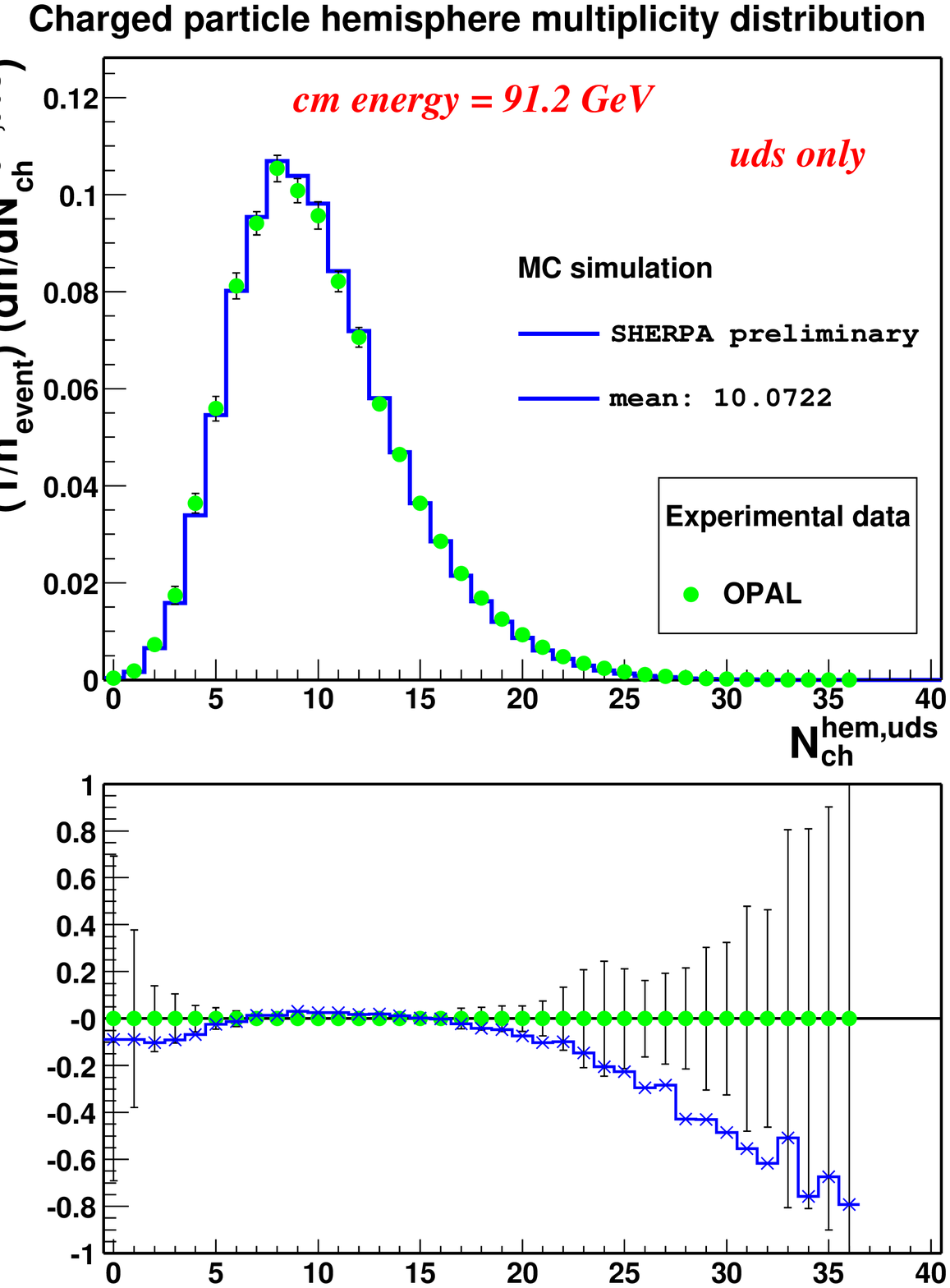,width=5.5cm}
        }&
      \mbox{\epsfig{file=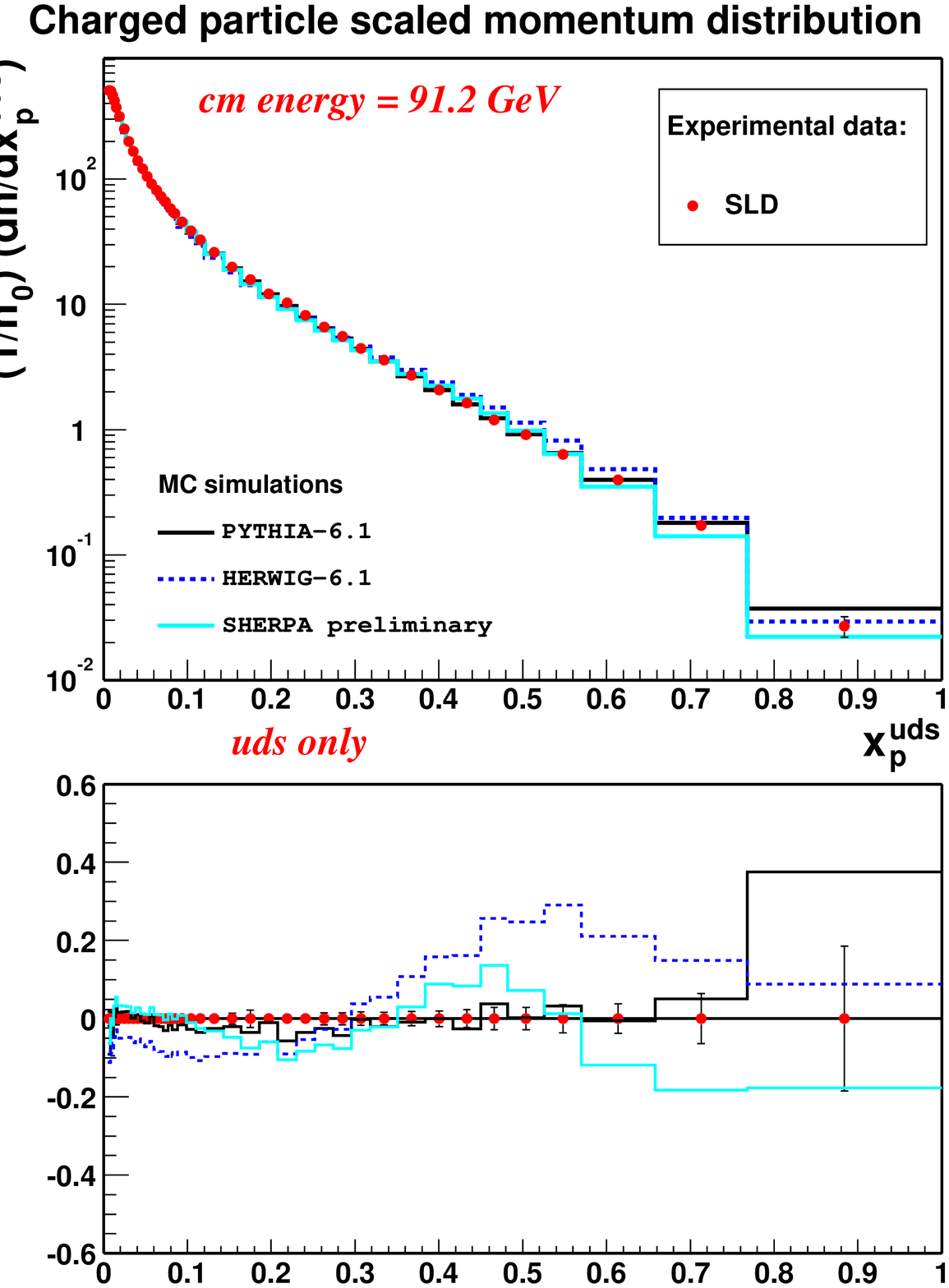,width=5.5cm}
       }
   \end{tabular}
 \end{center}
 \vspace{-1mm}
 \caption{Predictions for the hemisphere multiplicity distribution (left) and the 
scaled momentum distribution (right) of charged particles considering the 
light-quark sector only. Results are compared to OPAL and SLD data, respectively. }
 \label{fig:ChP_mult_xp}
\end{figure}

The model so far is restricted to the fragmentation of light quarks ($uds$) 
and gluons produced in $e^+e^-$ annihilation. Nevertheless various observables and 
distributions can be studied in order to validate the model.
\begin{table*}[t!]
  \vspace{0mm}\begin{center}\tiny
    \begin{tabular}{|c||c||c|c|c|}\hline    
      &&&&\\[-2mm]
      & $\langle\cal{N}^{\mr u \mr d \mr s}_{\mr{ch}}\rangle$
      & $\langle\cal{N}^{\mr u \mr d \mr s}_{\pi^{\pm}}\rangle$
      & $\langle\cal{N}^{\mr u \mr d \mr s}_{\it{K}^{\pm}}\rangle$
      & $\langle\cal{N}^{\mr u \mr d \mr s}_{\it{p},\bar{\it p}}\rangle$\\[2mm]\hline
      \multicolumn{5}{c}{}\\[-3mm]\hline
      &&&&\\
      {\tt PYTHIA-6.1($uds$)} &
      $19.84$ & $16.72$ & $2.010$ & $0.856$\\[2mm]\hline
      &&&&\\
      {\tt HERWIG-6.1($uds$)} &
      $18.86$ & $15.37$ & $1.693$ & $1.568$\\[2mm]\hline
      &&&&\\
      {\tt SHERPA$\alpha$} &
      $20.15$ & $16.83$ & $2.018$ & $1.047$\\[2mm]\hline
      \multicolumn{5}{c}{}\\[-3mm]
      \hline
      &&&&\\
      DELPHI \cite{Abreu:1998vq} 
      &
      $19.94\pm0.34$ &
      $16.84\pm0.87$ & $2.02\pm0.07$ & $1.07\pm0.05$\\[2mm]\hline
      &&&&\\
      SLD    \cite{Abe:2003iy} 
      &
      $20.048\pm0.316$ &
      $16.579\pm0.304$ & $2.000\pm0.068$ & $1.094\pm0.043$\\[2mm]\hline
    \end{tabular}
  \end{center}
  \vspace{-2mm}
  \caption{\label{tab:multipl}Overall mean charged-particle multiplicity, and production
           rates of charged pions, charged kaons and (anti)protons in $uds$\/ events
           at the $Z^0$-peak.}
\end{table*}
Figure \ref{fig:clumass} shows the mass distribution of the primary clusters at 
$e^+e^-$ collisions for different energies proving the universality of the approach.
Table \ref{tab:multipl} contains the mean multiplicities of $\pi^\pm$, $K^\pm$ and $p$, 
$\bar p$ in comparison with experimental $uds$ results and Pythia and Herwig. 
In Figure \ref{fig:ChP_mult_xp} the charged particle hemisphere multiplicity distribution 
and the charged particle scaled momentum prediction are compared to OPAL 
\cite{Ackerstaff:1997xg} and SLD \cite{Abe:2003iy} data, respectively, again taking into 
account $uds$ events only. 


\section{Conclusion}
The cluster-hadronization model developed for Sherpa proved to work successfully for 
$e^+e^-$ annihilation events into light-quark and gluon jets. First tests show a satisfactory 
agreement with experimental data. Some cluster model shortcomings, such as the too low 
charged-particle multiplicities, could be cured; and the spectrum of the scaled momentum could 
be improved. The model will soon be extended to cover heavy-quark hadronization and the fragmentation 
of beam remnants in hadron-hadron collisions.



\end{document}